# IoT service slicing and task offloading for edge computing

JaeYoung Hwang, Lionel Nkenyereye, NakMyoung Sung, JaeHo Kim, and JaeSeung Song, *Senior Member, IEEE*

*Abstract*—With the advancement of IoT technology, various domains such as smart factories, smart cities and smart cars use the IoT to provide value-added services. In addition, technologies such as MEC and network slicing provide another opportunity for the IoT to support more advanced and real-time services that could not have been previously supported. However, the simple integration of such technologies into the IoT does not take full advantage of MEC and network slicing or the reduction of latency and traffic prioritization, respectively. Therefore, there is a strong need for an efficient integration mechanism for IoT platforms to maximize the benefit of using such technologies. In this article, we introduce a novel architectural framework that enables the virtualization of an IoT platform with minimum functions to support specific IoT services and host the instance in an edge node, close to the end-user. As the instance provides its service at the edge node where the MEC node and network slice are located, the traffic for the end-user does not need to traverse back to the cloud. This architecture guarantees not only low latency but also efficient management of IoT services at the edge node. To show the feasibility of the proposed architecture, we conduct an experimental evaluation by comparing the transmission time of both IoT services running on the central cloud and those using sliced IoT functions in the edge gateway. The results show that the proposed architecture provides two times faster transmission time than that from the conventional cloud-based IoT platform.

## I. INTRODUCTION

In recent years, the Internet of Things (IoT) has been considered a technology that can solve various social problems [1]. In addition, various IoT devices such as home appliances, sensors, actuators, and drones can be connected and interact with each other. These devices are used in various domains such as home automation, smart grids, traffic management, and medical aid [2]. According to a report from McKinsey [3], it is expected that the IoT will have a total potential economic impact of $3.9 trillion to $11.1 trillion per year in 2025.

With the growth of IoT industries, various vertical IoT use cases have emerged, which require high bandwidth and reduced traffic delay to support real-time IoT services. In this regard, the existing solutions based on a centralized IoT platform in the cloud have difficulty satisfying such needs.

J. Hwang, L. Nkenyereye and J. Song (corresponding author) are with the Department of Information and Computer Security, Sejong University, Seoul, South Korea (e-mail: {forest62590@sju.ac.kr, nklionel@sejong.ac.kr, jssong@sejong.ac.kr).

N. Sung, J. Kim are with the Autonomous IoT Research Center, Korea Electronics Technology Institute (KETI), Gyeonggi-do, South Korea (e-mail: {nmsung@keti.re.kr, jhkim@keti.re.kr).

Additionally, processing the massive quantity of data generated by IoT devices is a significant burden on the central IoT platform [4]. To solve such problems, many research activities have considered new technologies as promising solutions, including multiaccess edge computing (MEC) to increase the real-time efficiency of data management and data transfer and artificial intelligence (AI) technologies [5].

5G network slicing is a technique that allows accommodating different quality of service requirements by exploiting a physical network infrastructure. To realize this concept, network function virtualization (NFV) and software defined networking (SDN), as key enablers, are driving the network paradigm shift [6], [7]. NFV and SDN allow the implementation of flexible and scalable network slices on top of a physical network infrastructure. When used together, MEC and NFV/SDN can provide solutions that are not possible when using these approaches individually. Their integration is expected to increase throughput as well as optimize the use and management of network resources.

However, as long as an application server is located in a cloud that is far away from an edge node, the throughput and latency saved using NFV/SDN and edge computing are limited. To fully take advantage of such technologies, we need to consider additional requirements that avoid network traffic traversing from an end device to an IoT service application in the cloud. For example, we can consider a cloud-based IoT platform that provides a set of common IoT service functions such as registration, data management and discovery. IoT devices or applications, regardless of which network technologies are used, have to connect such centralized IoT service platforms to store measured data or discover required information. If IoT devices are used in the smart home domain, which does not require high throughput and low latency, centralized IoT platforms work very well. However, different IoT use cases such as industrial and mission-critical IoT services require faster response times (e.g., less than a few milliseconds (ms)) and higher reliability. In such domains, even if the access and core networks are virtualized and deployed closer to the users using SDN/NFV, a message from an end IoT device or application must traverse to the IoT cloud platform. A response for the request should also make the same traversal from the IoT platform to the end IoT device, which cannot serve low-latency requirements, for example, within a few ms [8].

More specifically, as shown in Fig. 1 [9], even when network slicing occurs on physical network infrastructure, the round trip time (RTT) cannot be reduced tremendously because a request from an IoT user equipment (UE) must reach the cloud-based IoT platform. Therefore, a mechanism virtualizing IoT common service functions and deploying them to an edge node needs to be considered, and specific IoT services consisting of the IoT resources must also be delivered to the edge nodes. Although the IoT common service functions are supported to the edge nodes, if IoT services are still running in the cloud-based IoT platform, the traffics from the IoT UE have to traverse to this platform for the IoT services to be used. As a result, the following requirements can be identified.

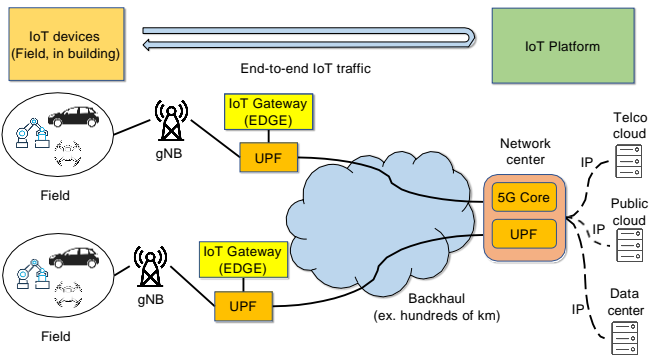

Fig. 1. End-to-end traffic where the IoT service platform is placed in the cloud

- IoT platforms running in the cloud should be able to create and manage multiple virtual IoT services containing only necessary common service functions.
- An edge node in a network slice should serve not only virtualized IoT service functions but also resources representing IoT data and services.
- To harmonize with the other network slicing technologies, the IoT platform should be able to exchange necessary information, such as the location of an edge node in which an MEC server is located, with the underlying network.

In this article, we propose a novel reference architecture to address the latency limitation of the current edge computing-based 5G network system from an IoT service perspective. The reference architecture introduces two core concepts to provide users with IoT services in proximity: ***IoT service slicing and IoT task offloading.*** IoT service slicing divides the holistic IoT platform by functionality to flexibly operate the IoT functions at edge nodes and adopts a software virtualization approach to make the IoT functions granular. IoT task offloading creates the IoT resources being operated in the cloud-based IoT platform to provide the same-level IoT services at the edge nodes. Therefore, with the adoption of these two concepts into the existing cloud-based IoT platform, IoT functions and services can be operated at a node close to users, and the traveling time of users' requests can be reduced. Accordingly, it is predicted that the IoT platform can provide users with more seamless and faster IoT services. In addition, to prove the feasibility of the newly defined reference architecture, a transmission time performance comparison between edge node-based IoT services and cloud-based IoT services is included. From the evaluation results, we can confirm that applying the proposed architecture to the edge node and the IoT platform can achieve better performance than just serving the IoT services by the cloud-based IoT platform.

The main contributions of our work are as follows:

- We propose a novel concept that virtualizes IoT common service functions to provide IoT services on the edge nodes that are close to the end-users. Our approach enhances the benefit of edge computing, as virtualized IoT services supporting a specific user or services can be deployed at the same edge node with a 5G network slice. By deploying both the IoT service slice and 5G network slice at the same edge node, it is possible to achieve high throughput and low latency.
- The implementation of the proposed architecture and performance evaluation showcase the feasibility and effectiveness of IoT service offloading through the experimental results, which show an average two times faster round trip time from the IoT device to the platform.

The rest of this article is organized as follows. Section II is included to compare the existing research. Section III explains background technologies that are the basis of IoT service slicing. Section IV introduces an overview of the proposed IoT service slicing architecture and how it can be incorporated with MEC technology. Section V evaluates the proposed architecture using a prototype implementation based on oneM2M IoT standards. In addition, in Section VI, we identify and discuss the remaining challenges that have to be addressed. Finally, in Section VII, the article finishes with a conclusion and discussion of future work.

## II. RELATED WORK

At present, to provide time-sensitive services with low latency, several edge computing-based studies have been conducted that show the feasibility of such approaches, with good results. More specifically, computation offloading is considered a promising approach to efficiently use the resources of both edges and clouds while reducing the burden of mobile devices. As one of the approaches to realize computation offloading, traditionally, most of the research is based on game theory and optimization theory. However, these factors are hindered by influencing factors such as being multidimensional, randomly uncertain, and time-varying [10]. Therefore, to address these issues, edge computing research based on AI has emerged.

In [11] based on the industrial IoT scenarios. In this regard, most edge computing research focuses on finding power-delay trade-offs and ignores the service accuracy supported by edge servers. Therefore, to solve the issues, the authors use the transfer learning approach to provide more improved service accuracy at the edge nodes by fine-tuning the pretrained network according to service type. As a result, this approach significantly reduces the computational complexity of the training phase and shows improved service accuracy. According to one study [12], deep reinforcement learning (DRL) with a sequence-to-sequence (S2S) neural network has been used to make a task-aware computation offloading model. Unlike the existing RL approach computation offloading research not considering the task dependency, this research considers the task dependency and models it by using a direct acyclic graph (DAC) as input data. By using a DAC, the sequential tasks are efficiently offloaded to the edge nodes or cloud. In contrast to the two aforementioned AI-based studies, another work [13] mentioned that a number of classical machine learning algorithms are not suitable for MEC because of massive computational demands. In addition, once the features of computation tasks including a number of mobile devices (MDs) and input data size change, newly trained models have to be produced by rerunning the whole scheme. Therefore, to solve these issues, a decision tree-based offloading scheme (DTOS) based on good adaptability was recently developed. In parallel, an approach to support the automated orchestration of end-to-end IoT services, the concept of IoT slicing based on microservices has been proposed [14]. The IoT slicing orchestrated by this approach can reduce the latency, and multitenant IoT solution coordination is possible despite the heterogeneity.

To conclude, various edge computing studies have been conducted to support the low-latency service, but to the best of our knowledge, reference architecture including MEC and IoT standards (e.g., oneM2M) and a mechanism for delivering the tasks built upon those standards to the edge nodes have not been adequately considered.

## III. BACKGROUND ON KEY CONCEPTS FOR IoT SERVICE SLICING AND TASK OFFLOADING

This section first describes network slicing enablers that are currently being actively researched as the network technologies for realizing tailored network services and then explains core technologies such as microservices and the oneM2M IoT standard used in this article to perform IoT service slicing.

### A. Network Slicing Enablers

Network slicing is used to create a logically divided network instance by dividing a physical network and can provide a dedicated network specialized for the service to users. Each network slice is guaranteed network resources such as virtualized server resources and virtualized network resources. In addition, each network slice is isolated from each other so that errors or failures caused in a specific network slice do not affect the communication of other network slices. To realize this concept, NFV and SDN have been considered enablers of network slicing.

NFV implements the network elements based on softwarization techniques and runs them on general-purpose servers (i.e., industry-standard servers, switches, and gateways) [15]. As a result, service providers could reduce capital and operational expenses (CAPEX/OPEX) since they do not need to launch network services that require space, power consumption, and massive effort to integrate the appliances. In addition, NFV abstracts the physical network resource and decouples the virtual network functions (VNFs) from the underlying infrastructure; thus, the hardware-independent lifecycle can be ensured [16].

In a traditional network system, routers use several network routing algorithms to deliver packets from the source destination to the target destination. In the process, there are two types of main operators. The control planes continually update the routing table, and data planes send the packet to the next hop. The problem is that if the network topology continually grows, the lists in the routing table are also increased. This situation leads to reduced routing performance since the control planes have to continually update the routing tables itself. Therefore, SDN, which has the entire view of the network topology, has been proposed to efficiently control the network flow, and the main goal of SDN is to decouple the control plane and data plane to efficiently manage the network flow. By using the above technologies, it is expected that network slicing is realized and that various vertical use cases are supported, such as smart homes, smart industries, and smart cities. However, simply slicing at the network infrastructure level is not enough to fully support IoT services.

### B. IoT function modulization based on microservice and virtualization

As a first step to provide IoT service slicing, modulization of the IoT platform is considered. At present, most IoT services are designed and implemented according to a centralized-cloud concept [17]. That is, all common IoT service functions including device management, registration, and discovery are deployed on a centralized cloud service. However, users need to receive different types of IoT services, and when deploying the modularized IoT functions to the edge, in terms of efficiently using the computational resources of edge nodes, all IoT functions do not need to be deployed at the edge nodes. In conclusion, it is important to divide the holistic IoT platform system into several small components and subsequently consider the appropriate methods to modularize the IoT platform. Currently, microservice architecture has been receiving much attention in industry fields as an agile delivery mechanism, and the International Data Corporation (IDC) expects that by 2021, 80% of applications being developed on cloud platforms will be microservices [18]. More specifically, microservice is a concept decomposing a monolithic system into a granular system, and through this approach, small applications based on the microservice approach can be deployed, scaled, and tested. Therefore, modulized IoT functions can not only independently deploy software but can also provide a more flexible and agile development environment by reducing centralized management.

Meanwhile, container-based architecture has emerged to reconstruct cloud computing, and it provides advantages such as good isolation, low overhead, and fast start-up time [19]. Therefore, along with the advantages of these two separate architectures, much research has been performed, and in this stream, to realize this architecture, Docker, which is a container-based open-source platform for the efficient use of physical resources, has been considered the most promising tool to implement microservices. Docker follows the container-based concept to efficiently manage the physical resources and abstract them to support a hardware-independent system. However, the increase in the number of Docker containers disturbs their management, and manual management is not recommended. Therefore, the docker system includes the

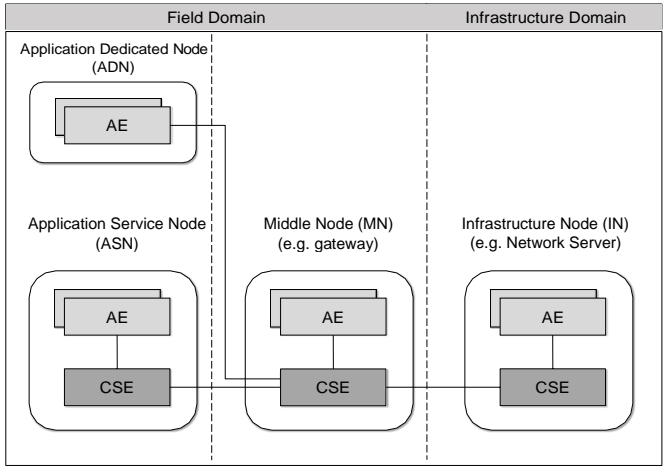

Fig. 2. oneM2M reference architecture

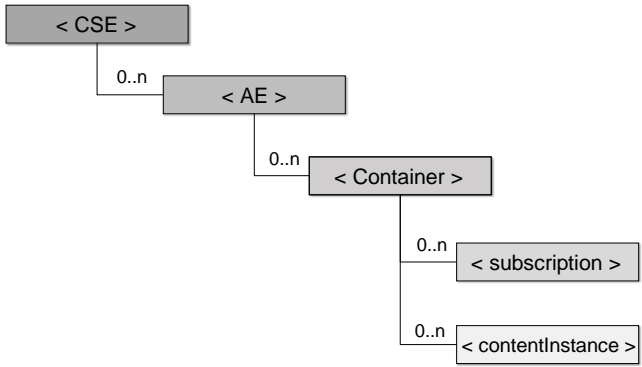

Fig. 3. oneM2M resource structure

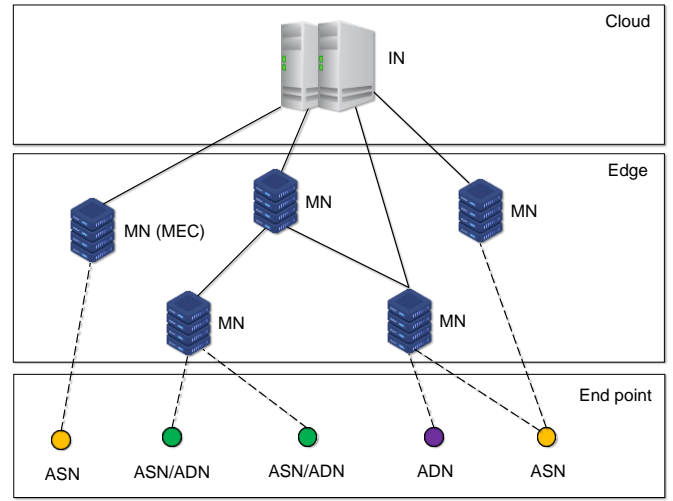

Fig. 4. oneM2M reference architecture with MEC concepts

Docker Swarm, which is the docker container orchestrator. Through this tool, Docker container creation, deletion, updating and other management functions can be easily executed [20]. In addition, Kubernetes, which is an open-source container orchestration system for managing operational resources, has been gaining much attention in 5G infrastructure as a container-based platform because this application has been acknowledged by a large number of users worldwide due to its scalability and efficiency of resource management [21].

### C. oneM2M standard overview

oneM2M is a global IoT standard body with seven leading information and communications technologies (ICT) standards development organizations: Association of Radio Industries and Businesses (ARIB) and Telecommunication Technology Committee (TTC) from Japan, Alliance for Telecommunications Industry Solutions (ATIS) and Telecommunications Industry Association (TIA) from the US, China Communications Standards Association (CCSA) from China, European Telecommunications Standards Institute (ETSI) from Europe, and Telecommunications Technology Association (TTA) from Korea. The aim of oneM2M is to develop a horizontal IoT service layer to mitigate fragmentation coming from a variety of IoT standards.

oneM2M is developing the reference architecture to assist a large number of IoT services such as smart factories, smart cities, and smart cars, and four functional entities called nodes are defined as described in Fig. 2 [22], [23]. The infrastructure node (IN) represents the IoT server in the reference architecture. The IoT gateway can be illustrated as a middle node (MN). IoT sensors and actuators are formed in the oneM2M environment as an application dedicated node (ADN) and application service node (ASN).

In addition, oneM2M adopts the resource-oriented architecture (ROA) model. That is, all IoT devices and data based on the oneM2M standard can be handled as resources based on a hierarchical structure [24], and each resource in the structure has an identifier to refer the resources. In this regard, several resource types have been defined, but in this article, basic resources to make the IoT services are considered [25], as described in Fig. 3. The `common service entity (CSE)` supports common service functions including registration, discovery, and data management. The `application entity (AE)` presents the various service logic types. The `<Container>` plays a role as data instances. The `<contentInstance>` has the real sensored value. The `<subscription>` resource can be created as a child resource that requires a subscription. Each `<subscription>` has notification policies that include which, when, and how notifications are sent [26]. The `<CSE>` is the root of all child resources in oneM2M, and functional nodes must comprise at least one `<CSE>` or `<AE>`. In the following section, the aforementioned technologies are combined to support IoT slicing and task offloading. More specifically, conceptual architecture and procedures regarding task offloading are illustrated.

### D. MEC from the oneM2M perspective

In the last few decades, mobile subscriptions have increased by up to 63%, and a variety of new vertical services are being provided including videos, gaming, smart cities, smart factories and so on. In addition, the advancement of ICT forces the existing centralized cloud system to require very good performance. Therefore, to overcome the challenges in terms of latency and guarantee high speed, the ETSI MEC industry specification group (ISG) has commenced developing standards including use cases, architecture, and application programming interfaces (APIs) to create an open environment and provide the vendor-neutral MEC framework [27], [28].

According to the current oneM2M research regarding Edge computing, one architecture model is proposed to integrate the MEC concept into oneM2M, and oneM2M sets the principle that the newly integrated edge computing concept has minimal impact on the existing oneM2M architecture [29]. In this context, the oneM2M node previously described can be presented as described in Fig. 4. The IN can be the cloud, the MN can be deployed either at the edge node or a normal gateway node, and the ADN and ASN can be the terminal devices. However, this conceptual architecture is not suitable for the full use of the current MEC architecture. Therefore, beyond simply applying the edge computing concept to oneM2M, the ETSI MEC ISG and oneM2M group are currently discussing how to provide an IoT-friendly environment by defining the MEC APIs for IoT systems.

## IV. IoT SERVICE SLICING AND TASK OFFLOADING

This section describes the conceptual architecture and procedures of how to achieve IoT service slicing by performing IoT function modulization and task offloading based on container technology and IoT standards. Detailed procedures for IoT service slicing and resource offloading are described based on MEC and oneM2M global standards [29], [30].

### A. IoT service slicing conceptual architecture

A cloud-based IoT platform where all common service functions are placed is generally enough to support IoT services such as smart home and smart building use cases that do not require very low latency. However, in other cases such as mission-critical IoT services based on less than a few ms, the existing IoT platform is not suitable since it cannot ensure a fast response time to satisfy mission-critical IoT service requirements. In general, conventional IoT platforms are located in the cloud, which is far from the devices and applications. In addition, as cloud IoT platforms provide all the required functions, if there exists a vast amount of data traffic from IoT devices, the resources in the cloud platform can be overloaded. Edge nodes can reduce such overload in the cloud by processing some or all data without transferring them to the cloud. Processing data at edge nodes requires the edge nodes to support relevant IoT service functions. Taking into consideration these requirements, we introduce a novel architecture for IoT service slicing and task offloading, as shown in Fig. 5.

First, we need a clear definition for IoT service slicing and task offloading as follows:

*IoT service slicing* is a concept that modularizes the common service functions of the IoT platform into small microservices that can be deployed at the edge nodes. IoT service slicing selectively collects only necessary IoT microservices and creates a virtual IoT platform instance to move the instance towards the edge nodes. This approach allows the IoT system to operate IoT services more flexibly and efficiently so that each IoT slice can be optimized for a specific IoT use case.

*IoT task offloading* is the transfer of IoT resources associated with the data to be processed at the edge nodes from the IoT cloud service platform. In the IoT platform following ROA, as all IoT services are represented in the form of resources in the cloud, it is essential to have necessary resources on the edge node to mimic the requested IoT service. In this context, a task is a set of resources created by specific users or administrators on the IoT platform to use or provide IoT services.

Based on the two definitions, the proposed architecture supports low-latency IoT services as follows. An IoT platform supporting IoT service slicing and offloading first takes a request from the IoT devices that need an IoT service with very low latency. As a next step, the platform checks which IoT common service functions are required to satisfy the request, and prepares an IoT slice that contains the required micro common IoT service functions. The platform then decides where to run the IoT slice based on the available MEC nodes around the IoT device. A selected MEC node hosts the instantiated IoT slice and performs resource offloading for the resources representing the requested IoT service.

The proposed architecture comprises six logical entities to support IoT service slicing and task offloading as follows:

- *IoT service common management layer (ISCL)* is a management layer that has entire views of IoT services currently operating in the cloud and an IoT service slicer that is responsible for slicing IoT services. Through this layer, administrators can understand the status of the currently running system through the logs.

- *IoT service slice manager (ISSM)* handles container-related requests from the ISCL. The ISSM is responsible for creating and storing the container images and has a collection of APIs for image management.

- *Common service function repository (CSFR)* is a repository to store container images for each IoT function. Basic IoT functions necessary for operating IoT services may be stored. In addition, according to specific requirements, additional components developed by the developer in accordance with the IoT platform standard may be stored in the CSFR. In this context, for example, Docker is used as a container technology. Docker uses a Docker image to run the container. Docker images are packages of application components required for each service, and since the configuration between each application is set in advance, they can be easily used in other new environments and are easy to manage and deploy [31].

- *IoT service slicer (ISS)* is the orchestrator for processing both IoT service slicing and task offloading. The ISS initiates container images of edge nodes according to the IoT service types. In addition, the ISS delivers tasks to a specific edge node to provide IoT services at a short distance from users.

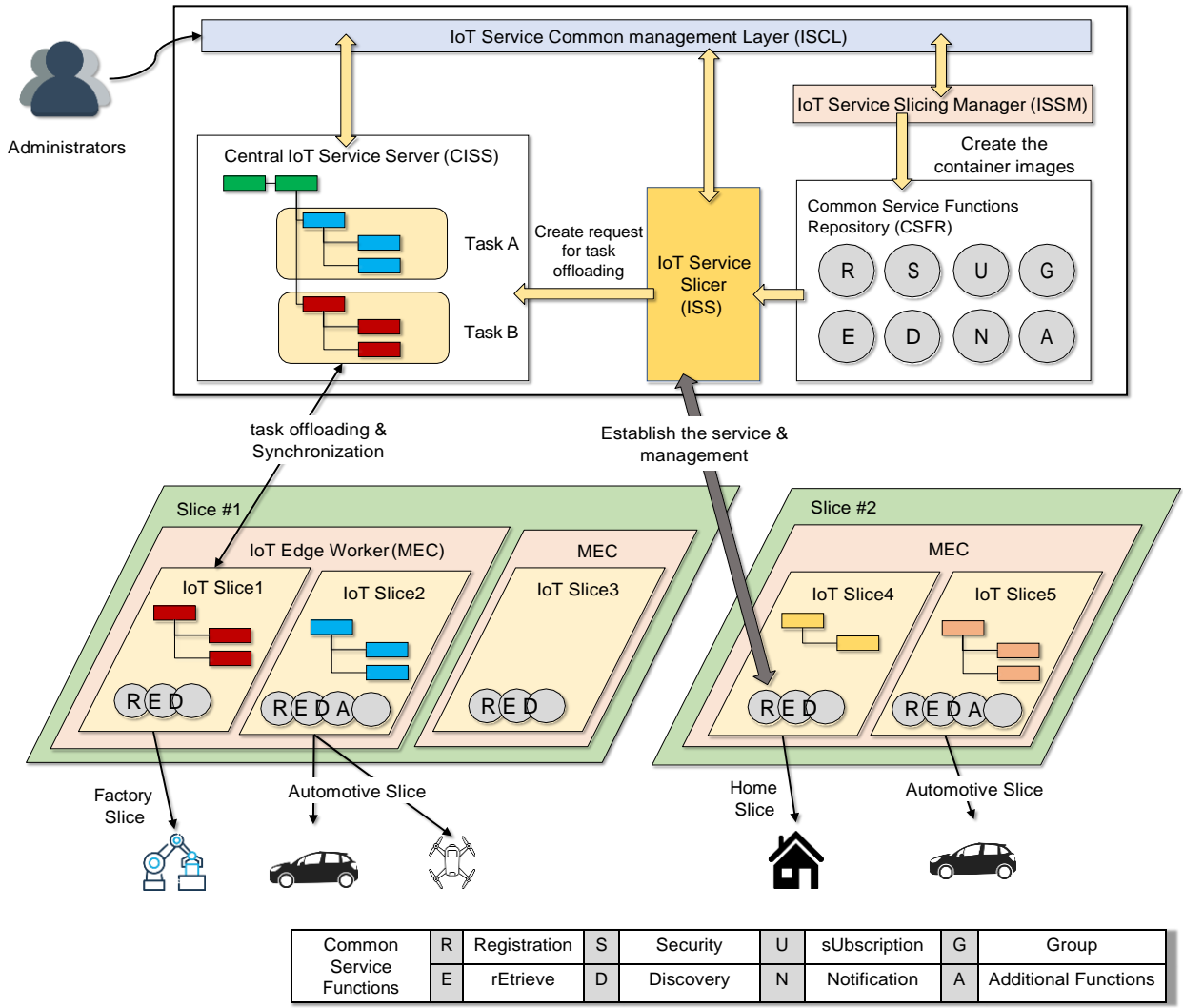

Fig. 5. Reference architecture describing IoT service slicing and task offloading

- *Central IoT service server (CISS)* can provide for all IoT functions in the cloud system and should have APIs between itself and each edge node to send commands and perform data synchronization. Therefore, the information processed at the edge node can be reflected in the IoT cloud to provide information through continuous updates. In addition, commands can be delivered through the cloud.

- *IoT edge worker* is the node that operates tailored IoT slices deployed by the ISS. The IoT edge worker is a *container-runner* running on MEC and can support container-based software such as Docker. In addition, at IoT edge workers, containerized IoT functions are deployed and operated to support the IoT services of each user.

### B. IoT service slicing based on microservices

In this article, we decide to use the microservice concept for our IoT service slicing approach. In general, deploying the IoT service platform that supports all common service functions at the edge nodes is not a relatively hard task if edge nodes have enough resources in terms of CPU and memory. However, providing the fine-grained graduality of IoT services based on microservices gains more advantages in terms of resource management because most edge nodes do not have enough memory and CPU processing power to serve an IoT service platform. First, from a resource management perspective, when the IoT system is terminated due to unexpected errors, edge nodes do not need to respawn the container having an IoT system, while if edge nodes adopt the microservices, they will just respawn the small-level microservices according to the IoT service type. For example, some simple IoT service use cases such as measuring the temperature do not require subscription-notification IoT functionalities. In this case, the edge node does not need to load IoT systems with subscription and notification functionalities in the hardware. As a second reason, currently, there are many IoT services such as smart cities, smart factories and smart homes, and these IoT services may have different requirements in terms of latency and computation resources. In this regard, IoT services with different requirements can be operated on the same edge nodes. If all IoT services are being managed by the monolithic IoT system, each IoT service will have the same service level regardless of their requirements. In this context, to guarantee the quality of experience (QoE) of each IoT slice, classifying which IoT service is requiring more resources and assigning the resources to them is important. For example, a smart home service does not have a time-sensitive condition as a requirement. In contrast, mission-critical services such as medical, drone, and smart car services definitely have time-sensitive requirements and may make use of more resources than smart home services. In summary, classifying the IoT service type and providing IoT functions based on microservices is one of the promising options for efficiently operating edge nodes.

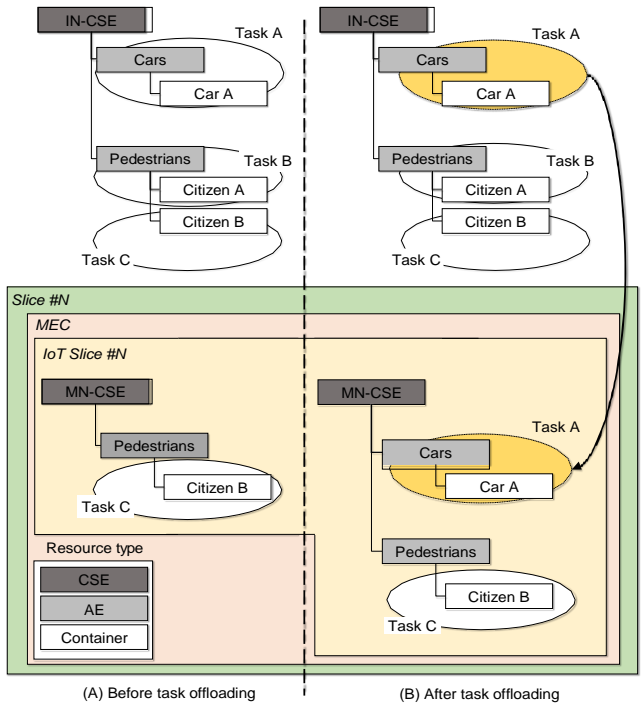

Fig. 6. IoT task offloading comparison

However, simply delivering the IoT functions to the edge nodes is not enough, and additional important procedures remain. IoT task offloading procedures that copy the IoT resources into the edge node must be considered.

### C. IoT service slicing and task offloading procedures

Another important requirement for realizing IoT service slicing is not only to containerize IoT platform functions implemented in the cloud but also to deliver actual tasks and data to an edge node where the containerized IoT functions are instantiated. In this context, a task is a set of resources representing physical things and containing data and meta-information on the IoT platform. Once the edge node is ready to serve a sliced IoT platform, relevant resources required to support the service need to be placed on the edge node by IoT task offloading. As shown in Fig. 6. (a), IoT physical entities are represented as a resource in the IoT platform.

For example, a location of Citizen B is represented as a re-source, `MN-CSE/Pedestrians/CitizenB/location`, in the edge node; the location is the `<contentInstance>` resource, and it does not appear in this figure. In addition, to support service for other tasks by the edge node, relevant resources should be placed and managed in the edge node.

Let us consider a road scenario that warns pedestrians crossing a crosswalk by sending the pedestrians' location to vehicles. In Fig. 6. (a), task A (IN-CSE/Cars/CarA/) and task B (IN-CSE/Pedestrians/CitizenA/) are running on an IoT platform in the cloud, and another task C (MN-CSE/Pedestrians/CitizenB/) is running at the edge node since Citizen B is walking in the smart city, providing edge-based IoT services. When assuming that task B is already offloaded and that Car A on the highway is entering the smart city if a request is made by the car to use an IoT service with a few ms requirements, the IoT platform in the cloud must deliver the task being operated to the edge node. That is, to operate the service on behalf of the central cloud, the central cloud must offload the tasks being performed in the cloud to the edge gateway. After going through these procedures, the resource structure of task offloading is changed to the structure shown in Fig. 6. (b). To recap, Car A can have faster IoT services than before owing to the shortening of transmission links. If the service is being served in the central cloud hundreds of kilometers away, it cannot meet the latency requirements for the service. Therefore, services such as road scenarios should be performed on the edge nodes near the services.

***Detailed procedures for service slicing and offloading:*** In these procedures, according to the IoT devices' request, first, IoT slicing is executed to provide the IoT functionalities to the edge nodes (steps 01-08). Next, IoT task offloading is performed to deliver the already operational IoT services in the central cloud to the edge nodes (steps 09-14). These procedures are materialized by using the components that we described above, and three main components, namely, ISSM, CISS, and ISS, are used.

01:IoT devices send a request to the MEC node to use edge-based IoT services. In this context, IoT devices can be mobile phones or many other devices based on the Internet such as drones or smart cars.

02:In the MEC specification [32], the ME App (MEC Application) that is running as virtual machines on top of the virtualization infrastructure is defined. In this article, the ME App can be an IoT slice with IoT function container instances. In addition, the IoT Slicing Handler, also an ME App, is defined to address IoT-related messages. More specifically, the IoT Slicing Handler is deployed in advance to act as an IoT message handler and holds information about the instantiated IoT functions per IoT slice that are currently being performed. Based on this information, the IoT Slicing Handler matches the request from the IoT device and checks whether required IoT functions for a requested IoT slice can be supported by currently running IoT slices. According to the matching result, the IoT Slicing Handler delivers a corresponding request message to the ISCL.

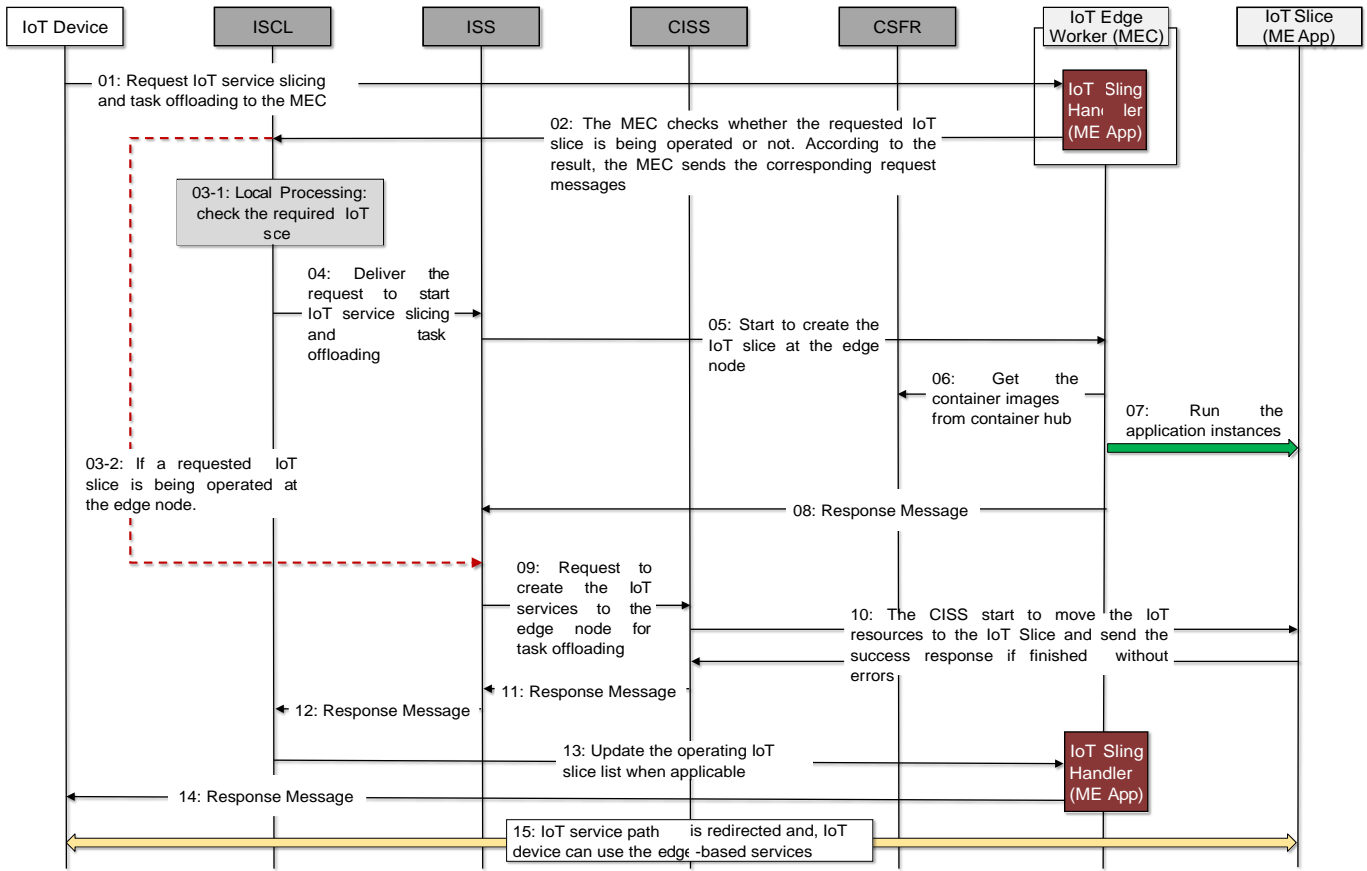

Fig. 7. IoT service slicing and task offloading initial procedures

03-04: The ISCL is responsible for structuring the request message for the ISS and sends the message with information such as IoT service information, necessary IoT functions, and target edge node information. If the ISCL receives a message that indicates the required IoT functions are already supported by one of currently running IoT service slices, it delivers the IoT task offloading start message to the ISS (step 03-2, Go to step 09) by omitting steps 4-8 (for performing IoT service slicing). In contrast, in case some of the requested IoT functions are not performed by one of IoT slices, the ISCL delivers an "IoT function generation and task offloading" message to the MEC (steps 03-1, 4).

05: The MEC defines data models and related procedures as well as various APIs based on REpresentational State Transfer (REST) to provide the edge computing services [33]. For instance, there is an API to check whether the MEC App instance operated in the MEC system is currently running (*MEC application support API* [34]) and a procedure to initiate or terminate a specific MEC App instance (*application instance life-cycle management* [35]). Therefore, the ISS exchanges information with the MEC to provide IoT services using these various MEC APIs and related procedures. The ISS sends the message to start creating a required IoT slice to perform requested IoT service functions that are not supported by the existing IoT slices on the edge node.

06-08: In this phase, if the IoT edge worker already has the container images regarding the IoT functions that are not currently running, then that service function is simply initiated. Otherwise, the IoT edge worker has to access to the CSFR to download the required IoT service container images (step 06). After downloading the required container images, IoT edge worker initiates the containers to provide IoT functions (step 07). After confirming that all containers have been deployed to the edge node, the deployment phase of the IoT function for IoT service slice is finished, and now, the edge node can be considered as having an operational "IoT slice (ME App)" (step 08). For reference, since the MEC standard is currently based on a virtual machine [36], it does not download images from a specific image hub when there are no images. That is, MEC previously stores related images before initiating an application through a procedure such as an *application package management procedure*. However, this article describes the procedures under the assumption that MEC is used on a container basis [35].

09-12: For IoT task offloading, an IoT task offloading request message is sent out into the CISS with the IoT service information of a user (step 09). The CISS delivers tasks and data that are running and stored in the cloud to edge nodes to provide the edge nodes with IoT tasks that are currently supported in the cloud (step 10). The edge that receives the information can create IoT tasks to an IoT slice running on the edge node itself to seemingly perform the IoT service within a short distance from the user. In addition, the additional API should be defined not only to synchronize the data between the IoT slice at the edge node and the cloud but also to deliver the command or exchange the data with the cloud. When all tasks running in the existing cloud are deployed to the IoT slice at the edge node, the edge gateway is finally ready to support the "proximity of IoT services" (step 11). When the ISS sends the response message to the ISCL, it includes the list of the newly initiated IoT functions of the IoT slice (step 12).

13-14: In these steps, if the IoT service is newly initiated or some IoT functions are newly initiated in the existing IoT service slices in steps 05-07, this information is reflected by the IoT slicing handler. Therefore, later, if the IoT device requests the same IoT service, the IoT slicing handler can aware that the existing IoT slices support required IoT service functions and will just request IoT task offloading.

15: After performing all the previous steps, IoT devices can directly receive required IoT services from the edge node providing IoT service slicing and task offloading.

*IoT task offloading synchronization:* As we mentioned in the CISS component description, an additional API is needed to synchronize the information between edge nodes and the central IoT platform. More specifically, although the data generated by the IoT devices are stored in the edge node, if these data are not synchronized with the database of the IoT platform existing in the cloud, IoT applications that are not using the IoT service slicing cannot use the latest updated information. Accordingly, all data stored or updated at the edge node must be synchronized with the data managed by the IoT platform existing in the cloud. To realize synchronization, the "subscription-notification" mechanism can be considered as a suitable method. The IoT platform subscribes to specific task offloaded resources of the edge node so that the IoT platform can receive the updated contents from the edge nodes whenever there exist any updates. If the data of the subscribed resource are changed (e.g., updated with a new value or deleted), the edge node generates a notification message including the information about the changed data and delivers it to the IoT platform. The IoT platform that receives the notification can analyze the messages and update the corresponding data in the cloud to be synchronized with the one in the edge node. For example, in the oneM2M IoT standard, there is a `<subscription>` resource, and this resource can be created under specific resources to check any changes of the resource as described in subsection III-C. When the status of the subscribed resource changes, the edge node checks the notification target uniform resource indicator (URI) and sends the changed information to the IoT platform. As a next step, the IoT platform updates the changed information to synchronize.

In addition, another synchronization approach can be considered. Among IoT service use cases, there might be a case such as measuring temperature in a building. In this case, data can be delivered to the IoT platform over a relatively long period since the temperature does not change often. In contrast, another use case such as smart cars and drones can generate the data rapidly and frequently, and they would produce massive quantities of data within a short time period. The former case can use the synchronization method mentioned earlier. However, the latter case can degrade the performance by continuously creating and transmitting a synchronization channel, which results in using many resources on both the edge node and cloud. Therefore, if the task offloaded resource at the edge node is updated, immediate synchronization better not be occurred in the latter case. Instead, when retrieve requests arrive at the cloud from the IoT applications, the cloud can redirect the service path to the edge node where an active IoT slice for the requested service is currently running. For instance, when a retrieve request for a specific resource is delivered to the cloud, after checking the existence of an IoT slice associated with the resource, the request is delivered to the edge node that runs the IoT slice. Then, the data most recently stored or updated are delivered to the IoT application. In this case, synchronization can be performed after the termination of the IoT service running in the edge node. As there exist many different scenarios, a proper synchronization mechanism can be different according to the operating environment conditions.

To demonstrate the advantages of IoT service slicing and task offloading in the subsequent section, a Docker-based IoT service slicing environment to perform function distribution and resource offloading is implemented.

## V. EXPERIMENTAL EVALUATION

To prove the advantages of our proposed IoT slicing and task offloading, this section uses oneM2M standards and implements the IoT platform based on microservices using Docker to perform container-based IoT function orchestration and IoT task offloading. Based on this implementation, we conduct a performance comparison of the latency between the IoT services running on the central cloud and those using the sliced IoT functions in the edge gateway.

### A. Experimental setup and Implementation

To evaluate the impact of the proposed IoT slicing concepts, Docker, which initiates the container images in the devices, is used to make the testing environment as described. In addition, to compare the performance before and after IoT service slicing and IoT task offloading, as shown in Fig. 8, two scenarios are conducted: a cloud-based IoT service (blue line) and edge-based IoT service (red dotted line). Figure 9 provides detailed information on how IoT microservices are deployed and operated at each edge node. In this scenario, it is assumed that the edge node is capable of operating a Docker-based service and that the container images for executing IoT microservices are already deployed. For example, container images such as Registration (R) and rEtrieve (E) are already deployed at an edge node, and accordingly, the edge node can initiate these images without accessing the Docker Repository Hub. IoT microservices to perform a specific IoT service can be arranged with Registration (62590), Retrieve (62591), Subscription (62592), and Notification (62593), and numbers in parentheses indicate the port number on which each microservice operates. In addition, each IoT microservice operates on the HyperText Transfer Protocol (HTTP). To focus on the performance and low memory, these IoT microservices are developed based on the Node.js Web framework, which is a server-side JavaScript environment [37]. The series of procedures for interacting with microservices are executed based on the Docker Remote Engine API, which provides an API for interacting with the Docker daemon [38]. The IoT task offloading procedure is the same as that described in section IV-C, and additional considerations for edge nodes and the IoT platform are how to synchronize the data change between them. Therefore, after IoT task offloading, as described in Fig. 3, the `<subscripton>` resource is created under each `<container>` resource to check the status change such as data update. If data are updated at the edge node, a notification message is automatically configured and sent to the IoT platform to perform the data synchronization.

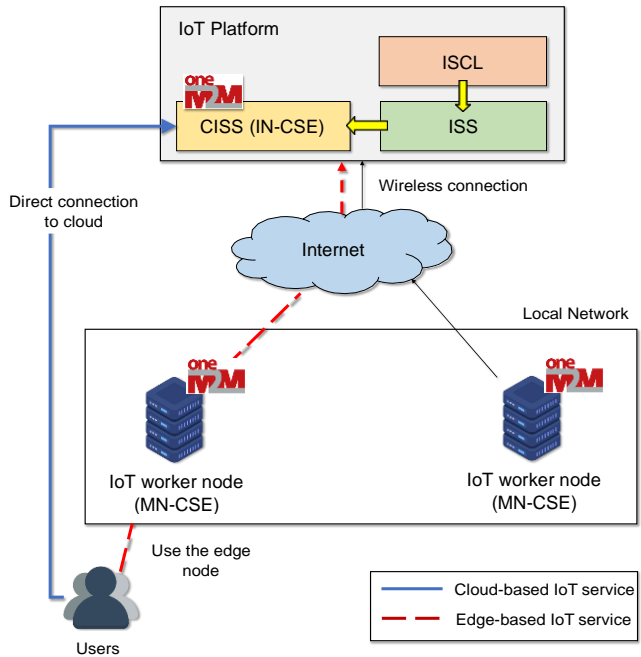

Fig. 8. IoT service slicing evaluation environment

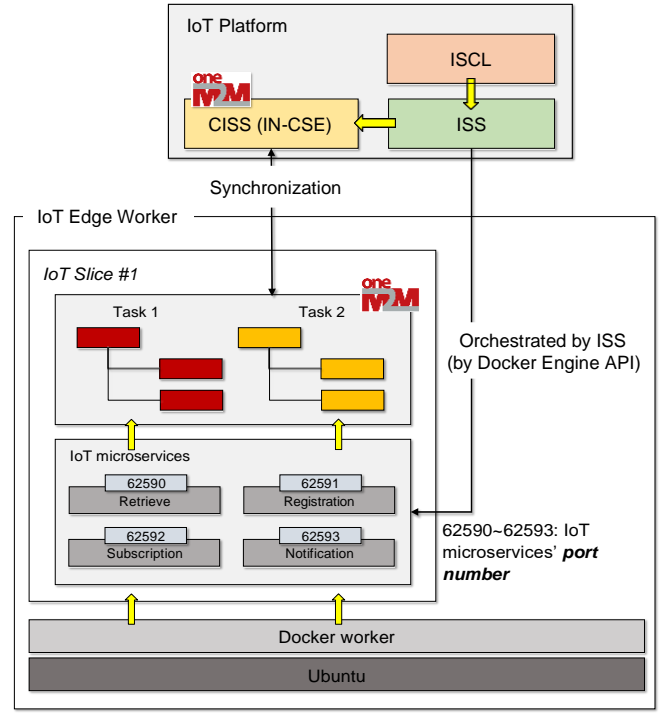

Fig. 9. IoT service slicing microservices on an edge node

### B. Evaluation

The RTT of both scenarios, a cloud-based IoT service and edge-based IoT service, is measured by using Apache JMeter. Apache JMeter is open-source software that tests functional behavior and measures performance. This software was first designed to test the web application, but through the extension of the test function, it can test the database and query, FTP server and so on [39]. As an evaluation method, three performance matrices are used when IoT service slicing and task offloading are finally performed: data update time through the creation of `<contentInstance>` resource and data retrieval time through the access of `<contentInstance>` resource. In addition, the average performance was presented by measuring the time from requesting to use the edge service to receiving the final completion response.

The data generated in the cloud and the edge node were assumed to continuously change the location of the device so that IoT devices will continually send the changed location as a form of `<contentInstance>` resources. The size of the `<contentInstance>` resource used in this comparison is approximately 400 bytes, and in order to measure the performance of `<contentInstance>` resource creation of both cloud and edge nodes, a request for creation is delivered 60 times, and the result is shown in Fig. 10. When data were delivered to the cloud, the performance distribution of the cloud was averaged at 8.5 ms. In addition, when the edge node takes the registration request from the IoT devices, on average, the performance distribution is approximately 6.1 ms. Therefore, in this context, approximately 2.0 ms of performance improvement is expected when edge nodes are brought in.

In addition, checking the data gathering time is one of the important parts of performance evaluation, allowing both the time regarding data generation to be measured and the IoT service-related data to be quickly used. Therefore, the most recently generated `<contentInstance>` resource including location information is used to evaluate the impact of the edge node. At first, the cloud used approximately 67.42 ms on average to respond to the clients. In contrast, the edge node spends the time delivering the response at approximately 37.32 ms on average. In summary, by using the edge node, it is likely that users can obtain more seamless IoT services.

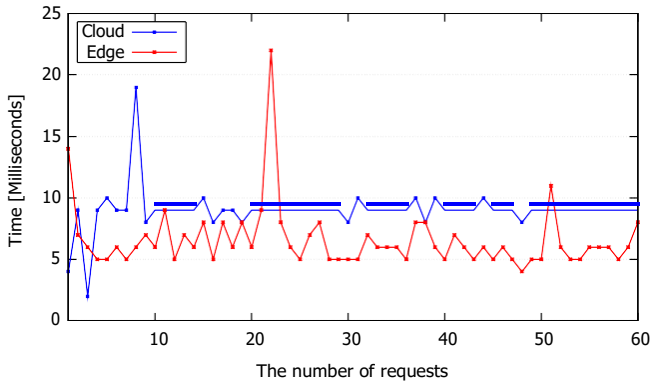

Fig. 10. Data generation time comparison between cloud and edge modes

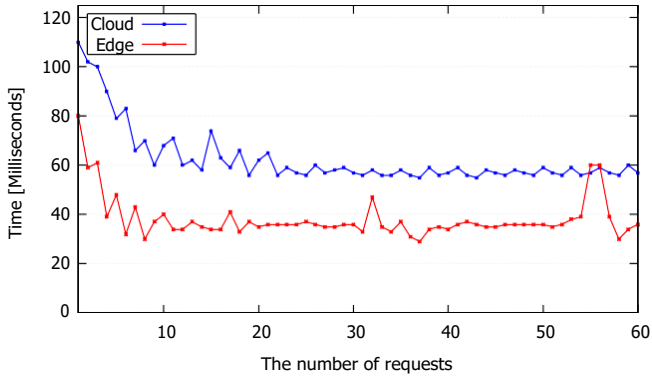

Fig. 11. Data retrieval time comparison between cloud and edge modes

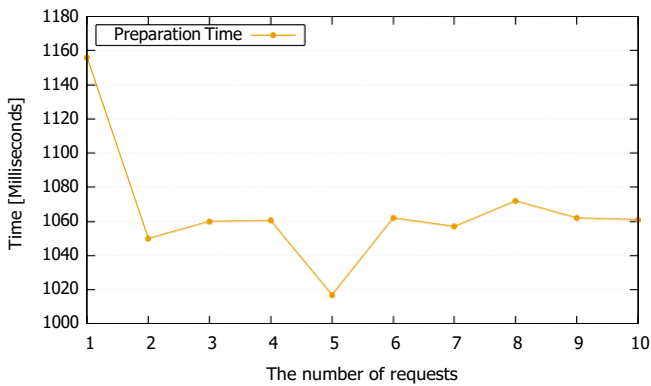

Fig. 12. IoT service slicing and IoT task offloading preparation time

By measuring the IoT service slicing and IoT task offloading time, it is possible to measure the service preparation time for the user to receive the edge node-based IoT services. To measure the corresponding time, a total of 10 measurements were performed. As shown in the graph in Figure. 12, the time of 1.047 ms was measured on average. This is the time when the image is already included in the edge node to operate the Docker service. If the Docker images are not already placed in the edge node, to download the image, the edge node needs to download the image from the Docker hub, and it might take a long time. For example, the size of the Node.js web framework image used in this evaluation is approximately 400 megabytes (MB), and the Python web framework such as Django is approximately 150 MB. Therefore, the IoT microservices required for supporting each IoT service do not change significantly unless a function update is required, such as adding a new function, so it may be better to store IoT function images in advance.

## VI. LESSONS AND LEARNED

Although we proposed the IoT slicing and task offloading approaches to cope with the current issues coming from the existing cloud-based IoT system from the IoT perspective, there are still problems that have to be addressed. In this section, the main challenges are identified and discussed.

***IoT slice standard interoperability:*** In most of the network slicing literature, the following problems are mentioned as the main challenges: service migration, security, and orchestration. First, regarding service migration, users will continually move with their mobile devices or IoT devices from one site to other sites. Therefore, to provide seamless services and a reasonable quality of service (QoS) level, services currently being used in the source edge node have to be migrated to the destination edge node according to the users' movement [40]. In terms of security from the network slicing perspective, essentially, network slicing is based on the virtualization of physical resources. In addition, security and privacy are considered major barriers to applying 5G networks [41], which is because virtualized network functions are deployed and shared on various service provider infrastructures. According to one study [42], orchestration is the continuing process for selecting the network resources to fulfill the client's requests regarding the service demands in an optimal way, and orchestrators as software enablers are responsible for automating the creation, monitoring and deployment of the network resources [43]. However, orchestrating different vendors' software or hardware network resources is not an easy task.

These three challenges can also be the main challenges to IoT slicing and task offloading. At present, the interoperability issue has been gaining much attention in IoT industries, as this challenge has to be solved first due to a large number of heterogeneous IoT protocols and standards. Interoperability must be guaranteed to deliver IoT services regardless of what IoT specifications or protocols companies are using [44]–[46]. This issue can also affect the IoT slicing and offloading architecture in terms of container management, service migration, and security. Each IoT standard has different security policies and data management policies. Accordingly, without concrete consensus among IoT standards, it is impossible to provide seamless IoT services to users. For example, suppose that edge node A uses the oneM2M IoT standard, while another edge node B is being operated on different IoT standards in the same infrastructure. When administrators need to migrate the IoT service from node A to B, if there is no consensus on how to authorize the incoming request and to convert the oneM2M standard to another standard, it is impossible to orchestrate all nodes efficiently, even though all nodes are being operated in the same infrastructure. Therefore, the IoT industry is actively researching interoperability to solve these problems, and in order to provide entire end-to-end IoT services to the mobile network, these issues have to be addressed.

***IoT slice performance isolation:*** There can be multiple IoT slices running on the same edge node. That means IoT slices for smart homes and buildings, which do not require much real-time processing, can be operated simultaneously together with slices for time-sensitive IoT services. To ensure performance of the running IoT services, each IoT slice must be guaranteed to use required resources whenever they need. For instance, Docker systems can explicitly set the amount of resources that each container can use when creating a container. In the evaluation, we do not consider performance isolation, but policies regarding limit resource quotas for each container to avoid disrupting the behavior of hosts and other containers must be considered and applied.

***Modularization of IoT platform:*** When evaluating the IoT service slicing and offloading scenario considered in Section V, we newly developed oneM2M-based microservices to perform IoT service slicing. Currently, there are two widely used oneM2M-based open-source IoT plat- forms, *Mobius* (http://developers.iotocean.org/) and *OM2M* (https://www.eclipse.org/om2m/), which have already been applied to various IoT industry fields including smart cities, smart factories, and so on. However, since not only the two aforementioned IoT platforms but also most of the current IoT platforms are not developed with initial consideration of microservices, it is difficult to modularize and virtualize IoT common service functions from IoT platforms. Therefore, in order to support IoT service slicing and offloading at the edge nodes, a microservice architecture should be considered from the beginning as a main design feature to be implemented.

***IoT slice security and trust using blockchain:*** Blockchain works based on a decentralized public ledger that can store records of transactions and complements the existing centralized ledger approaches that exhibit low efficiency arising from the bottleneck, single point of failure and security attacks [47]. All nodes on the network store data in a block structure logically connected to each other based on the hash value. These data blocks are copied and shared to be distributed to the entire network along with the blockchain system to protect the network system from cyber attacks and system failure. Due to the nature of the blockchains, it is evaluated as a technology that reduces the possibility of data forgery and alternations and prevents the act of producing illegal data for malicious purposes [48], [49].

At present, when operating an edge computing-based network system, there might be several security issues. First, an MEC infrastructure has heterogeneous features that cross organizational boundaries. That is, because MEC involves multiple different domains and MEC instances are managed by multiple organizations, there is a need to transparently and securely collaborate with other servers or applications [50], [51]. In addition, in the edge computing network environment, the data including privacy information must be partially or completely outsourced to other edge nodes or cloud data centers. Therefore, ownership and control of data are separated, which will lead to data loss or illegal data operations such as replication and publishing. Accordingly, data integrity cannot be guaranteed [52] in such environment. As a result, once such crafted data are transferred to other edge nodes or cloud IoT platforms, it could cause successively severe problems. Finally, during message transmission, adversaries could attack to disable the communication links by congesting the network or to sniff the network data flow. Thus, configuration data written by network administrators must be trustworthy and validated. However, this requirement is very challenging due to the high dynamism and openness of the edge computing environment. In this regard, the blockchain connecting MEC instances together with IoT platforms as a blockchain network could be considered a promising solution to increase the trust of slicing technologies [53].

In this regard, IoT slicing can also have advantages when it is applied with the blockchain technology. For example, currently, oneM2M-based IoT platforms support a resource called `<accessControlPolicy>` to control access right, and `<accessControlPolicy>` allows to store the authentication key value generated internally in the IoT platform to be used for device authentication. When a user or administrator tries to perform an operation to a resource representing an IoT application, the IoT platform checks the received authentication key value matches. However, this approach is not sufficient to verify the reliability of data exchanged among distributed edge nodes. For instance, if an IoT device using IoT services from the edge nodes, data on the IoT device, distributed edges and the central IoT platform should be the same based on sychronization mechanisms. When data of edge node A needs to be migrated to edge node B, it is not clear whether an authentication is needed and how to authenticate the data. In addition, the transmitted data are continuously delivered in the form of a data chain of several distributed edge nodes. In such situation, it is important to check whether the data shared or transferred among distributed edge nodes are the same without any modification. Additionally, when performing synchronization of edge node data to the cloud, it is essential to check whether it is data in trust as there are possibilities to mitigate data during the synchronization. Therefore, by connecting associated entities using the blockchain, it is possible to check the data integrity.

As shown in Fig. 13, by incorporating blockchain technologies into edge nodes where data are processed locally, the system can support more reliable security and trust among distributed edge nodes. In addition, as edge nodes and relevant instances, i.e., IoT devices, and the cloud are connected within the same blockchain network, when synchronizing edge node data to the cloud, to achieve the necessary agreement on single data, the information can be trusted by a consensus mechanism used in blockchain systems. Using the proposed blockchain enabled IoT slicing architecture, various information and data can be stored and shared to distributed blockchain entities. For example, the following transactions and queries can be supported by the proposed architecture:

- The oneM2M platform can query the blockchain and

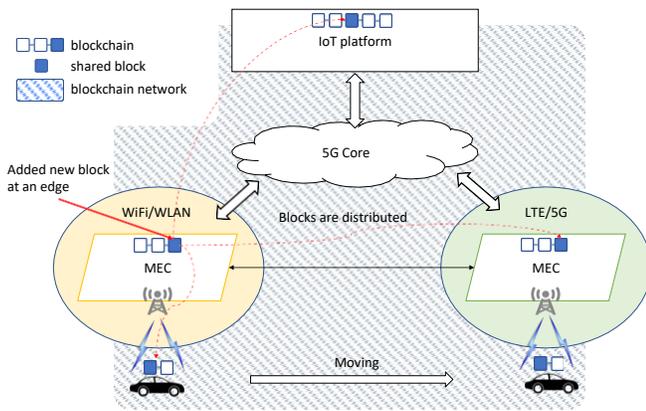

Fig. 13. MEC enabled blockchain for IoT slicing and offloading

determine that there is a set of IoT slices
- Request available MEC edge nodes that support IoT slice and offloading based on service level agreement between IoT service providers and network providers
- Inform the IoT service platform that IoT slice and offloading was being placed
- Share the updated or newly created data to blockchain nodes in the blockchain network

## VII. CONCLUSION

With the advancement of IoT technology, various domains use the IoT to provide value-added services. In addition, technologies such as MEC and network slicing provide another opportunity for the IoT to support more advanced and real-time services. However, providing real-time services via conventional IoT service platforms in the cloud shows various limitations since the user traffic needs to be delivered to an IoT platform in the cloud center even if MEC nodes host network slices. Therefore, in this paper, a reference architecture applying IoT slicing and task offloading mechanisms was presented, and the results based on experimental evaluation proved the advantages of the proposed reference architecture in terms of reducing the latency. As lessons learned from our experience, the main challenges, i.e., interoperability, security, and trust, were identified and discussed in future work. To conclude, this research shows the feasibility of the proposed architecture, especially in providing time-sensitive IoT services. In the future, we plan to enhance the system to support the dynamic management of containers by using AI technologies.

## ACKNOWLEDGMENT

This work was supported by an Institute for Information & Communications Technology Promotion (IITP) grant funded by the Korean government (MSIT) (No. 2018-0-01456, AutoMaTa: Autonomous Management framework based on artificial intelligent Technology for adaptive and disposable IoT)